\documentclass[10pt,letterpaper,twocolumn]{article} 

\usepackage{ol2}
\usepackage[draft]{hyperref}
\usepackage{amsmath}
\usepackage{cite}

\begin{document}

\twocolumn[ 

\title{High-power broadband laser source tunable from 3.0~$\mu$m~to~4.4~$\mu$m based on a femtosecond Yb:fiber oscillator}


\author{Tyler W. Neely$^{1,*}$ Todd A. Johnson,$^{1,2}$ and Scott A. Diddams$^{1,3}$}

\address{
$^1$National Institute of Standards and Technology, Time and Frequency Division,\\
Mail Stop 847, 325 Broadway, Boulder, Colorado 80305, USA\\
$^{2}$Present address: Saint John's University, Collegeville,
MN 56321, USA\\
$^3$email: scott.diddams@nist.gov\\
$^*$Corresponding author: tneely@nist.gov
}

\begin{abstract}We describe a tunable broadband mid-infrared laser source based on difference-frequency mixing of a 100 MHz femtosecond Yb:fiber laser oscillator and a Raman-shifted soliton generated with the same laser. The resulting light is tunable over 3.0~$\mu$m~to~4.4~$\mu$m, with a FWHM bandwidth of 170~nm and maximum average output power up to 125~mW. The noise and coherence properties of this source are also investigated and described.\end{abstract}


 ] 

High repetition-rate mid-infrared (MIR) femtosecond lasers and frequency combs present useful light sources for absorption spectroscopy, combining broad bandwidth with high spectral resolution and brightness~~\cite{Adler:10,Vodopyanov:11}. Such sub-picosecond sources have typically targeted the region from 2~$\mu$m~to~20~$\mu$m, where many molecules exhibit strong fundamental transitions. In particular, several schemes using single-pass difference frequency generation (DFG) have been developed and are attractive due to their relative simplicity and the benefit of passive carrier-envelope offset (CEO) frequency stabilization~\cite{Baltuska2002}.

DFG schemes based on Er:fiber oscillators have been successful; up to 1.1~mW of average power has been achieved in the range of 3.2~$\mu$m~to~4.8~$\mu$m~\cite{Erny2007}; 3~mW average power has been achieved at 3~$\mu$m~\cite{Maddaloni2006}; 1.5~$\mu$W was achieved in the region of 9.7~$\mu$m~to~14.9~$\mu$m~\cite{Winters2010}; and 100~$\mu$W of average power has been produced in the range of 5~$\mu$m~to~12~$\mu$m~\cite{Gambetta:08}. The power levels achieved thus far are however not comparable with watt-level powers available from optical parametric oscillators (OPOs)~\cite{Adler2009}. We present here a 100 MHz DFG-based system that for the first time achieves output power levels that are competitive with OPO techniques. At the same time, we examine the noise properties of our source and demonstrate that the nonlinear Raman shifting we employ can suffer from excess amplitude and phase noise, with the result of reduced pulse-to-pulse coherence in the MIR light. These results should be relevant for related DFG approaches to broadband MIR generation~\cite{Winters2010,Gambetta:08}.

We generate a MIR frequency comb with straightforward difference frequency generation (DFG) between the amplified output of a 100~MHz repetition rate mode-locked Yb:fiber laser and a Raman-shifted soliton~\cite{Mitschke1986}. The oscillator consists of both fiber and free space elements~\cite{Nugent-Glandorf:11,Zhou:08} and generates a spectrum centered at 1.04~$\mu$m with 140~mW average power. The oscillator operates in the similariton regime~\cite{Ilday:04} and produces strongly chirped pulses with a duration of about 1.5~ps. Additional chirp is introduced with an 8~m length of positive second-order, negative third-order dispersion fiber. Amplification occurs in a double-clad Yb:fiber reverse-pumped with up to 8~W of multimode light centered around 976~nm. Average output power and pulse duration after isolation and compression are 2.4~W and 130~fs.\\
\begin{figure}[t!]
\centerline{\includegraphics[width=8.3cm]{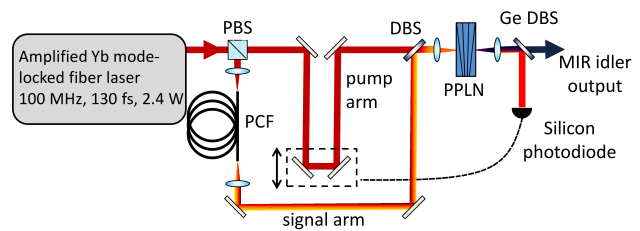}}
\caption{(color online) Experimental setup. PBS polarizing beam splitting cube, PCF photonic crystal fiber, DBS dichroic beam splitter, PPLN fan-out periodically-poled lithium niobate, Ge DBS anti-reflection coated germanium filter.}
\vspace{-1em}
\label{fig1}
\end{figure}
A simplified diagram of the experimental setup is shown in Fig.~1. Part of the amplifier output is coupled with an efficiency of $\sim$ 43~\% into a photonic crystal fiber (PCF) (1~m, 3.2~$\mu$m core, 945~nm zero-dispersion wavelength), primarily generating a red-shifted Raman soliton through asymmetrical broadening. Adjustment of both the input polarization and launch power into the PCF results in a signal pulse tunable from~1.10~$\mu$m~to~1.63~$\mu$m, as shown in Fig.~2(a). DFG occurs in a fan-out periodically poled MgO-doped lithium niobate (MgO:PPLN) crystal with poling period varying from 21~$\mu$m~to~35~$\mu$m and corresponding poled length varying from 1.2~mm~to~2~mm, allowing for continuous tuning of the phase-matching condition. The crystal temperature is held at $100~^{\circ}{\rm C}$ to ensure that photorefractive damage is avoided. The signal is combined with the original amplified pump centered at 1.04~$\mu$m, resulting in an idler wave in the MIR ($\lambda_i^{-1} = \lambda_p^{-1} - \lambda_s^{-1}$). Temporal overlap of the pulses is accomplished by introducing extra optical delay into the pump path, compensating for the dispersion-induced delay in the PCF. The pulse overlap is monitored using coincidental non-phase-matched sum-frequency generation (SFG) between the soliton and pump pulses ($\lambda_{sum}^{-1} = \lambda_p^{-1} + \lambda_s^{-1}$); the resulting sum light is centered around 630~nm and is measured with a silicon photodiode. A 5.8~kHz dither of the Yb:fiber amplifier pump diode current causes a propagation delay dither at the output of the PCF and a subsequent dither in SFG power. Lock-in detection of the SFG power generates an error signal that drives both a moving mirror pair at sub-hertz bandwidth (Fig.~1) and the Yb:fiber amplifier pump diodes at bandwidths of a few hundred hertz.

The resulting MIR idler is thus tunable from 3.0~$\mu$m~to~4.4~$\mu$m as shown in Fig.~2(b), with a FWHM spectral width of $\sim$ 170~nm and maximum average power of 128~mW at $\sim$~3.2~$\mu$m. A Gaussian fit to this spectrum gives a peak spectral power density of $0.67\pm0.01$~mW/nm. As shown in the inset of Fig.~2(b) for the 3.2~$\mu$m tuning point, the output power is approximately linear with pump power above a given threshold, with a slope of $93\pm3$~mW/W. This suggests that high-output MIR powers may be possible with pump-power scaling; 1~W of average idler power might be expected with 11.2~W of pump power, comparable to the output of PPLN-based OPO systems~\cite{Adler2009}.
\begin{figure}[t]
\centerline{\includegraphics[width=8.3cm,height=9cm]{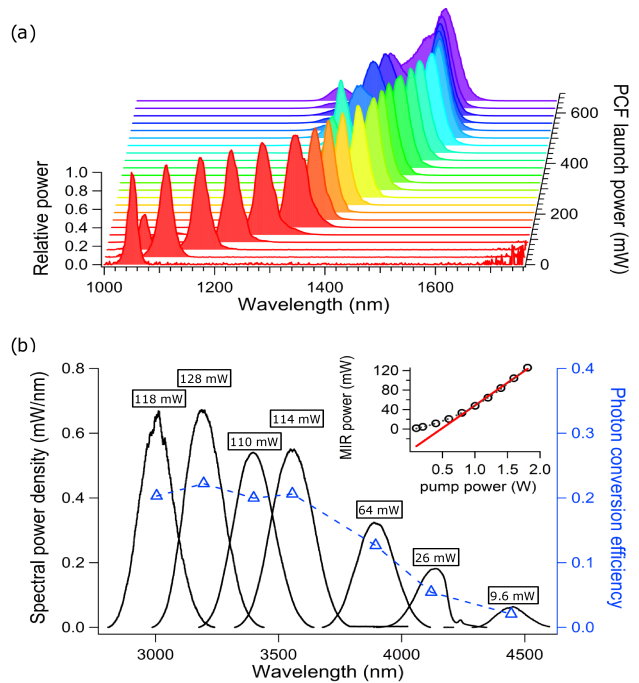}}
\caption{(color online) (a) Soliton shift as a function of launch power into the PCF, with each spectrum individually normalized. (b) MIR idler spectra, and conversion efficiency relative to pump photons. The total average power is also noted above each spectrum. The absorption feature at 4.4~$\mu$m is due to air-path absorption of CO$_2$ before the monochromator. (inset): Average idler power as function of pump power at the $\sim$~3.2~$\mu$m tuning point. }
\vspace{-2em}
\label{fig2}
\end{figure}
\begin{figure}[t]
\centerline{\includegraphics[width=8.3cm,height=8.8cm]{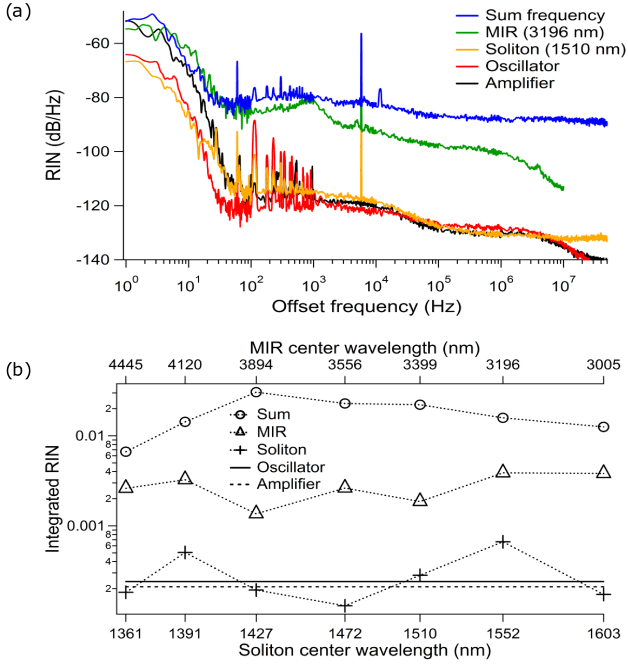}}
\caption{(color online) (a) Relative intensity noise (RIN) for the MIR tuning point of 3.196~$\mu$m. Measurement of the MIR RIN is limited to 10~MHz due to photodiode bandwidth. (b) Integrated RIN (10~Hz~-~10~MHz).}
\vspace{-1em}
\label{fig3}
\end{figure}

The decreased photon conversion efficiency at longer wavelengths is not understood, although it is consistent with results from similar systems based on Er:fiber oscillators~\cite{Erny2007,Gambetta:08}. Throughout the tuning range, the soliton pulse lengths varied between 97~fs and 117~fs, pulse powers varied between 33~mW and 51~mW, and spot size at the PPLN varied between $\omega_0=78$~$\mu$m and  $\omega_0=91$~$\mu$m. Taken together, the variation in these parameters is however insufficient to explain the measured variation in output MIR power, and we have been unable to tie this behavior to an experimental parameter.

The utility of this source for producing a MIR comb with passive CEO stabilization depends on both on the phase (PM) and amplitude (AM) noise of the generated light. It is known that asymmetrical broadening, as employed here, can lead to excessive AM and PM noise~\cite{Dudley2006}. Additionally, AM noise on the amplifier output couples via dispersion to PM noise on the soliton, which couples back to AM noise on the MIR light. We have attempted to quantify the origin and magnitude of noise sources by heterodyne and relative intensity noise (RIN) measurements at each stage of the system.

RIN measurements of the system are shown in Fig.~3. The RIN of the MIR light (0.3~\% integrated, 10~Hz~to~10~MHz) is found to be about a factor of 5~to~10 greater than the RIN on the soliton. An estimated RIN-induced timing jitter of the soliton of 4.5~fs (10~Hz to 10~MHz) is obtained from the product of the integrated amplifier RIN (0.00021), the average PCF launch power (350~mW), the measured PCF dispersion (70~fs/nm), and the average power-dependent soliton wavelength shift (1~nm/mW). However, the impact of this amount of jitter on the cross-correlation signal between a Gaussian-approximated pump (130~fs) and soliton (100~fs) amounts to MIR power variation of only 0.04~\%, which is insufficient to account for the measured results in the range of 0.3~\%. We thus attribute the observed increase in MIR RIN to excess intrinsic timing noise on the soliton that arises in the nonlinear spectral broadening process.  Measurements of the SFG RIN are additionally included in Fig.~3. The SFG also depends on pulse timing overlap, but its RIN measurement is not limited by photodiode bandwidth.  The reason for the measured increase in RIN for the SFG (compared to the MIR RIN) is not clear at this point.

As an additional characterization we measured the heterodyne beat between the soliton and a tunable CW diode laser with a balanced InGaAs detector. Fig.~4(a) shows the results of the measurement of the free-running beat notes at the soliton center wavelength of 1.36~$\mu$m, with a 300~kHz integration bandwidth. As the soliton shift is increased, the contrast of the beat note decreases to zero, as shown in Fig.~4(b), suggesting poor pulse-to-pulse coherence. The MIR spectrum was also investigated by doubling the light generated at 3.1~$\mu$m in a AgGaS$_2$ crystal, resulting in $\sim$ 100~$\mu$W of light at 1.55~$\mu$m. The heterodyne between this doubled light and the CW laser is shown in Fig.~4(c), and is compared with similarly doubled light from a MgO:PPLN-based OPO in the lab. Relatively high beat-note contrast is seen in the OPO, and no beat note is observed in the DFG system. Since the amplifier output generates a beat note with a Nd:YAG laser at 1064~nm (20~dB contrast in a 300~kHz bandwidth with similar light levels), we conclude that the excess noise is due to the soliton.

Although limitations are evident in the Raman soliton generation process, the simplicity, power and brightness of this source are still useful for ongoing broadband spectroscopic studies, including high-resolution trace gas spectroscopy~\cite{Johnson:11}. The system may be improved with careful choice of non-linear fiber and/or pulse parameters. In particular, recently developed suspended-core fibers have shown promise for highly coherent supercontinuum generation in the tunable range utilized in this experiment; such fibers may allow the development of high-power comb sources based on these techniques~\cite{Ruehl2011}.

We thank M. Hirano, Y. Kobayashi, I. Hartl, L. Nugent-Glandorf, and F. Adler for their contributions and helpful comments. This work was funded by NIST and the United States Department of Homeland Security's Science and Technology Directorate. As a contribution of NIST, an agency of the US government, this work is not subject to copyright in the US.

\begin{figure}[t]
\centerline{\includegraphics[width=8.3cm,height=8cm]{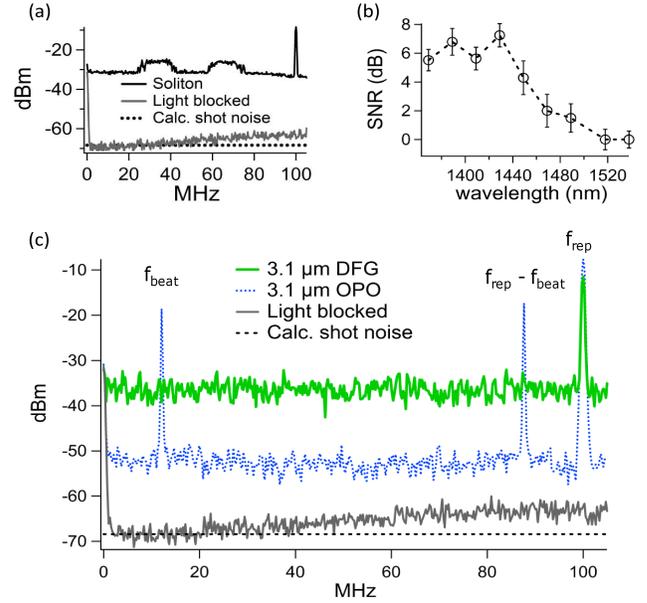}}
\caption{(color online) Coherence properties of the system. (a) Free-running beat note between the 1.36~$\mu$m soliton and CW laser with the max-hold function enabled over several seconds of measurement; the spread in beat-note frequency is due to variation in the unlocked repetition rate of the Yb:fiber oscillator. (b) Beat note signal-to-noise ratio across the soliton tuning range. (c) MIR beat notes.}
\vspace{-1em}
\label{fig4}
\end{figure}

\end{document}